\documentclass[pra,showpacs,twocolumn]{revtex4}
\newcommand{\uk}{u_{\mathbf{k}}}
\newcommand{\vk}{v_{\mathbf{k}}}
\newcommand{\uq}{u_{\mathbf{k}'}}
\newcommand{\vq}{v_{\mathbf{k}'}}

\newcommand{\epsk}{\epsilon_{\mathbf{k}}}

\newcommand{\Vkq}{V_{\mathbf{k},\mathbf{k}'}}
\newcommand{\kvec}{\mathbf{k}}
\newcommand{\qvec}{\mathbf{k'}}
\newcommand{\rvec}{\mathbf{r}}

\usepackage{graphicx}

\begin{document}

\title{Resonance effects on the crossover of bosonic to fermionic
superfluidity}

\author{S. De Palo}

\affiliation{Dip. di Fisica and INFM, Universit\`a di Roma ``La
Sapienza'', Italy}

\author{M.~L. Chiofalo}

\affiliation{INFM and Classe di Scienze, Scuola Normale Superiore,
Pisa, Italy} 

\affiliation{Istituto Nazionale per la Fisica Nucleare}

\author{M.~J. Holland} 

\affiliation{CRS-BEC INFM and Dipartimento di Fisica, Universit\`a di
Trento, via Sommarive 14, I-38050 Povo, Italy}

\affiliation{JILA and Department of Physics, University of Colorado at
Boulder, Boulder CO 80309-0440 USA}

\author{S.~J.~J.~M.~F. Kokkelmans}

\affiliation{Eindhoven University of Technology, Eindhoven, The
Netherlands}

\begin{abstract}
Feshbach scattering resonances are being utilized in atomic gases to
explore the entire crossover region from
a Bose-Einstein Condensation (BEC) of composite bosons to
a Bardeen-Cooper-Schrieffer (BCS) of Cooper pairs.
Several theoretical descriptions of the crossover
have been developed based on an assumption that the
fermionic interactions are dependent only on the value of a single
microscopic parameter, the scattering length for the interaction of
fermion particles. Such a picture is not universal, however, and is
only applicable to describe a system with an energetically broad
Feshbach resonance. In the more general case in which narrow Feshbach
resonances are included in the discussion, one must consider how the
energy dependence of the scattering phase shift affects the physical
properties of the system. We develop a theoretical framework which allows 
for a tuning of the
scattering phase shift and its energy dependence, whose parameters can
be fixed from realistic scattering solutions of the atomic physics. We
show that BCS-like nonlocal solutions may build up in conditions of
resonance scattering, depending on the effective range of the
interactions. 
\end{abstract}


\maketitle

\section*{Introduction}

Fermion pairing is a fundamental concept in the manifestation of
non-trivial ground states in condensed-matter physics.  Condensation
of composite fermions, viewed as bound states of an electron and an
even number of vortices, is a possible explanation of Integer and
Fractional Quantum Hall Effects in the highly degenerate
two-dimensional electron gas, the vortices corresponding to the
fractionally charged quasi-particles of Laughlin's
theory~\cite{QHE}. Exciton formation and possibly Bose-Einstein
condensation~\cite{BECEXC} in semiconductor structures is one more
example~\cite{EHLIQ}.  Cooper pairing resulting from correlations in
momentum space is the mechanism determining the BCS-type
superconductivity in metallic compounds~\cite{bcs}, while strong
correlations~\cite{Rice} and real-space pairing with a short coherence
length characterize the high-temperature superconductors, where pair
correlations manifest in the opening of a pseudogap well before the
superconducting transition~\cite{HTSC}.

The achievement of quantum degeneracy in atomic Fermi gases of
$^{40}$K and $^{6}$Li after cooling in dipolar optical
traps~\cite{Debbie,Hulet,ENS,LENS,Duke,MIT}, and of accurate control
of the interactions by means of Feshbach resonances~\cite{feshbach},
has made these concepts accessible in atomic-physics experiments. This
has initiated an exploration of an intriguing system which should
reveal the predicted resonance superfluidity
~\cite{holland,timmermans,servaasPRA} and the nature of BCS pairing
with large attractive interaction~\cite{bcsatoms}.

\begin{figure}[ht]
\includegraphics[width=3.5in]{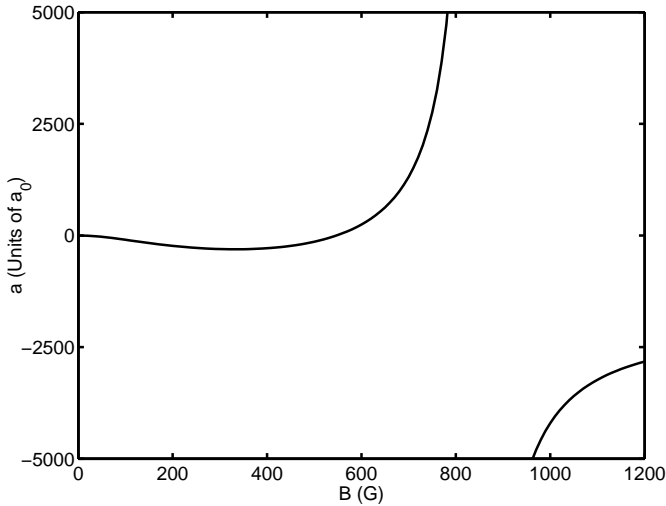}
\caption{The broad Feshbach resonance in $^6$Li~\cite{servaasPRA}. The
scattering length $a$ is shown as a function of magnetic field, for
the $(f,m_f)=(1/2,-1/2)$ and $(1/2,1/2)$ mixed spin channel.}
\label{fig:a}
\end{figure}

In these experiments, the magnetic field is changed to tune the
position of a bound state in a closed channel in the interatomic
potential with respect to the threshold of zero scattering energy. The
zero energy is defined by the asymptotic value of an open channel
potential in which the colliding atoms enter. At exact resonance, the
bound state connects with the zero energy scattering solution and the
scattering length is infinity. Variations of the magnetic field~$B$
with respect to the resonance value $B_0$ can be converted into an
effective detuning $\nu=(B-B_0)\Delta\mu^{\rm mag}$, with
$\Delta\mu^{\rm mag}$ the relative magnetic moment between the
closed-channel bound state and the open-channel threshold.  As
displayed in Fig.~\ref{fig:a} for the case of $^6$Li, positive
(negative) detunings with respect to the resonance correspond to
effectively attractive (repulsive) interactions resulting in negative
(positive) scattering lengths $a$. The Feshbach mechanism evidently
involves a separation of energy and length scales between the
background and the resonant behavior, the former driven by the value
$a_{bg}$ at very large $\nu$, and the latter dictated by the position
and width $\Delta\nu=\Delta B\Delta\mu^{\rm mag}$ of the
resonance. We point out that certain physical systems have large
values for the background scattering due to the presence of a
potential or shape resonance in the open channel, and a multiple
resonance model should be developed in that case~\cite{servaasabg}.

The side of the resonance corresponding to repulsive interactions
($a>0$) gives rise to a rich and complex quantum system.  Here, the
formation of weakly bound molecular states of two fermions has been
obtained ~\cite{debbiecontrol,ketterlecontrol,salomoncontrol} and
their Bose-Einstein condensation (BEC)
~\cite{debbiemolecular,grimmmolecular,ketterlemolecular,ENSmolecular}
has been observed.  On the other side of the Feshbach resonance
($a<0$) where the interactions are attractive, experimental
measurements are being made on the superfluid paired state that is
likely associated with a strong-coupling version of the
Bardeen-Cooper-Schrieffer (BCS) theory.

By continuous tuning of the interactions across the Feshbach
resonance, the whole crossover region from the relatively simple
limits of a BEC of preformed pairs and a BCS-like superfluid state is
accessible in the experiment. There are many open questions in the
understanding of the region in between, including the relevant
correlations in the system and the nature of the order
parameter~\cite{bosonfermion}.  Several theoretical descriptions of
the BCS-BEC crossover in Fermi gases with Feshbach
resonances~\cite{griffin,milstein,strinati,chicago} have been
developed, inspired by the early seminal works by
Leggett~\cite{leggett} and Nozi\`eres and Schmitt-Rink~\cite{NSR}, and
subsequent developments in the context of high-T$_c$
superconductivity. 

The crossover theories can be divided in two
general classes: studies in which the interactions are parametrized by
the scattering length~\cite{crossover}, and those that explicitly
include the presence of molecular states of two fermions and the
hybridization of these paired states with their unpaired
counterpart~\cite{RaFL}. Both approaches share the idea that the
formation of Cooper pairs and their condensation to the coherent
superfluid state do not occur at the same time. There are formal
connections based on either application of the Hubbard-Stratonovich
transformation to the functional integral~\cite{popov}, or to the
dressing of the open and closed channel solutions~\cite{bosonfermion}
which demonstrate the equivalence of the approaches when the resonance
state has a sufficiently short lifetime.

Two relevant questions arise in the quantitative description of the
crossover behavior in atomic gases. First, an adequate many-body
theory providing the crossover order parameter must be able to recover
the results of the four-fermion scattering problem entering the
Feshbach physics~\cite{shlyap}.  Second, the energy dependence of the
scattering phase shift may be important over the energy range of the
Fermi energy ~\cite{bruun}, and thus the width of the resonance
becomes a relevant parameter of the theory in addition to $a$. In the
unitarity limit $|a|\rightarrow\infty$, where the scattering length is
meaningless, the universal behavior of the thermodynamic properties
emerges only in the case of broad resonances~\cite{universal}, and otherwise the
specific details of the resonance will enter.  All
these aspects pose stringent conditions to the choice of suited
theoretical approaches. Along these lines, numerical methods
would be a very useful guideline to assess the validity of
approximate schemes and their underlying physics.

In this work we focus on the issue of the energy dependence of the
scattering phase shift in the unitarity limit, anticipating that the
general case of a BCS-BEC crossover driven by two independent
parameters (the detuning and width of the resonance) will lead to an
intriguing and nontrivial phase diagram.  Along these lines, we
develop a model that is able to interpolate between the two limits of
a broad resonance, where the scattering length resulting from
low-energy resonance renormalization completely encapsulates the
interaction behavior, and of a ``high-quality'' resonance in a broad
Fermi sea, where large values of scattering energies need to be
sampled.  We model an interaction potential that is composed of a
short-range attractive well followed by a barrier structure, where
scattering length and resonance width can be independently tuned. We
use such a model potential to show that a nonlocal BCS-like superfluid
ground state may emerge under non trivial conditions. The results are
indicative at this stage, as they are obtained within a mean-field
approach that is not necessarily valid in the regime of strong
interactions. However, it is a necessary precursor to perform such
calculations to guide the more complicated Quantum Monte Carlo (QMC)
simulations, that are under way.

The structure of the paper is as follows. We first develop an explicit
microscopic model based on a potential-well and a potential-barrier to
encapsulate the essential physics of the Feshbach resonance. We then
derive the nonlocal BCS equations for this system, assuming a closure
of the hierarchy of many-body correlations at the BCS level. Finally,
the BCS equations are self-consistently solved and the results
discussed in view of their application to the development of a more
detailed QMC simulation needed to capture the effects of many-particle
correlations.

\section*{A well-barrier model of the Feshbach resonance}

\begin{figure}[ht]
\includegraphics[width=3.2in]{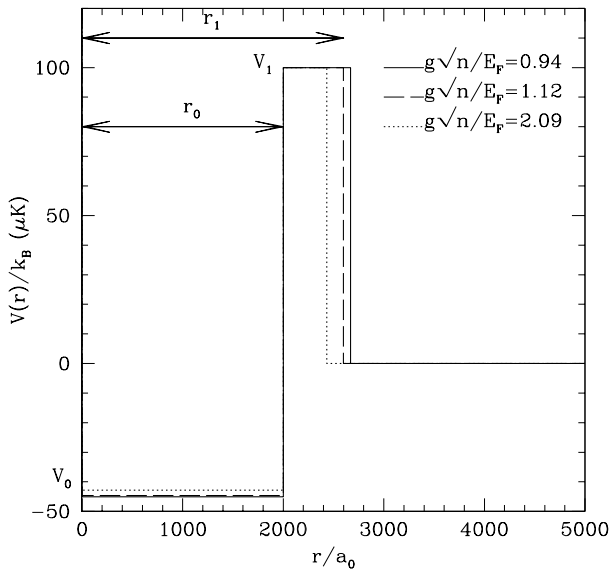}
\caption{A well-barrier model that allows for the independent tuning
of the scattering length $a$ and of the resonance width $g$. The 
well can support resonant states with a width dependent on the
tunneling through the potential barrier. The situations shown
correspond to the values used in subsequent simulations which all have
$a=5000a_0$, where $a_0$ is the Bohr radius. Different cases are
represented by the solid, dashed and dotted lines correspond to
increasing values of the resonance width $g$ (cases 1, 4, and 8 in
Table~\ref{tab:cases}).}
\label{fig:wb}
\end{figure}

We model the presence of a Feshbach resonance by the interaction
potential of the form displayed in Fig.~\ref{fig:wb},
\begin{equation}
V(r)=
\left\{\begin{array}{ll}
-V_0\qquad&r<r_0\\
V_1&r_0<r<r_1\\
0&\mbox{otherwise}
\end{array}
\right.
\label{eqVr}
\end{equation}
that is characterized by an attractive well with depth $V_0$ and width
$r_0$ and a barrier with height $V_1$ and width $r_1-r_0\equiv r_w$.
This well-barrier model allows us to incorporate
the essential energy-dependence of the scattering physics into a single-channel
scattering scheme. 

The condition for diluteness of the gas is that the range of the
interatomic potential should be much less than the interparticle
spacing, a condition that we assume to be fulfilled through this
paper. Explicitly this requires us to satisfy at all times
$nr_0^3\ll1$, where $n$ is the particle density. It is important to
emphasize that this is distinct from the condition $na^3\ll1$
associated with the convergence of the perturbation theory of the gas
in the contact scattering approximation. In the vicinity of the
resonance, we may have the situation in which $|a|>r_0$ so that we
simultaneously have a system which is dilute in the sense that
$nr_0^3\ll1$ but unitarity limited in the sense that $na^3>1$.  We can
furthermore distinguish the two different regimes of narrow and broad
resonance based on a comparison of the Fermi energy
$E_F=\hbar^2k_F^2/2m$ with $k_F=(3\pi^2n)^{1/3}$ with the resonance
width $\Delta\nu$. The resonance width $\Delta\nu$ can be expressed in
terms of the matrix element $g$ for the coupling between the closed
and open channels as $\Delta\nu=g\sqrt{n}$, that enters the
resonance-superfluidity Hamiltonian, as pointed out in
Ref.~\cite{pair}.

The parameters of the model potential can in principle be adjusted to
reproduce the scattering properties of an atomic sample, as {\it
e.g.}\/ they are determined from collision experiments or from full
coupled-channel calculations~\cite{servaasPRA}. We instead perform
here a study over a set of parameters chosen in order to illustrate
the important physical behavior of the system. We fix the scattering
length to a large and positive value, $a=5000\; a_0$, with $a_0$ the
Bohr radius. We ensure this is large compared to the interparticle
spacing as determined from the density which we take as $n=1.054
\times 10^{14}\,\mbox{cm}^{-3}$. The large positive scattering length
corresponds to the region of unitarity limited behavior considered
just on the BEC-side of the resonance. With these fixed constraints we
vary the resonance width crossing the full region of broad to narrow
values. To this aim, we first set the range $r_0$ of the potential to
a value that is small enough to satisfy the condition $nr_0^3\ll
1$. This is the case for $r_0=2000\; a_0$ giving $nr_0^3=0.125$. We
intentionally choose this value here so that it is not too small which
will allow us later to see on a single energy scale both the
scattering and many-body effects. Then, we solve for $V_1$ (the
barrier height) and $r_1$ (the barrier width) which are not
independent and can be tuned in order to change the tunneling rate
through the barrier and therefore $\Delta\nu$.

This requires solving the scattering problem for $V(r)$. The two-body
scattering function $\Psi(r)$ is given in the three regions by
\begin{equation}
\Psi_w(r)=
\left\{
\begin{array}{ll}
\Psi_w(r)\quad&r< r_0\\
\Psi_b(r)&
r_0< r< r_1\\ 
\Psi_f(r)&{\rm otherwise}
\end{array}\right.,
\label{eqPsi2b}
\end{equation}
with
\begin{eqnarray}
r\Psi_w(r)&=&\sin(k_wr) \nonumber\\
r\Psi_b(r)&=&A_1\exp[-k_b(r-r_0)]+B_1\exp[k_b(r-r_0)]\nonumber\\
r\Psi_f(r)&=&A_2\sin(kr)+B_2\cos(kr)
\end{eqnarray}
and 
\begin{eqnarray}
k_w&=&\sqrt{k^2+mV_0/\hbar^2},\nonumber\\ 
k_b&=&\sqrt{mV_1/\hbar^2-k^2}.
\end{eqnarray}
The values of $A_{1,2}$ and $B_{1,2}$ are determined after imposing
the usual boundary conditions $\Psi_w(r_0)=\Psi_b(r_0)$,
$\Psi'_w(r_0)=\Psi'_b(r_0)$, $\Psi_b(r_1)=\Psi_f(r_1)$, and
$\Psi'_b(r_1)=\Psi'_f(r_1)$, {\em i.e.}\/ ensuring the continuity of
the wave function and of its derivative at the boundaries $r_0$ and
$r_1$. The two-body $T$-matrix can then be extracted from the
scattering solution
\begin{equation}
T(k)=\frac{4\pi \hbar^2 B_2}{mk(iB_2-A_2)}.
\label{eqTk}
\end{equation}

We can now proceed to link the parameters of the potential with the
parameters of the many-body theory we wish to solve. The first
parameter, the scattering length $a$ is determined simply from its
definition applied to Eq.~(\ref{eqTk}), 
\begin{equation}
\frac{4\pi\hbar^2a}m=\lim_{k\to 0}T(k) .
\label{eqinva}
\end{equation}
The width of the resonance is related physically to a matrix element
between a continuum scattering state and a high-quality resonant state
in the inner well. In practice, such a matrix element is determined by
the tunneling rate through the barrier. In order to make the link
explicit we present the usual relation between the $T$ and
$S$-matrices in scattering theory
\begin{equation}
T(k)=\frac{2\pi \hbar ^{2}i}{mk}\left[ S(k)-1\right]
\label{eqTf}
\end{equation}
and give the Feshbach form for the $S$-matrix in the one resonance
parametrization
\begin{equation}
S(k)=e^{-2ika_{\rm bg}}\left[ 1-\frac{2ik|g|^{2}}{-\frac{4\pi
        \hbar ^{2} }{m}(\nu-\frac{\hbar ^{2}k^{2}}{m }) +ik|g|^{2}}\right] \; .
\label{eqSf}
\end{equation}
Solving Eqs.~(\ref{eqTf}) and (\ref{eqSf}) for $g$ gives
\begin{equation}
|g|^2=-\frac{8\pi\hbar^4}{m^2R_{\rm eff}}\;
\label{eqReff}
\end{equation}
where we have removed the background $a_{\rm bg}=0$ and the effective
range is by definition $R_{\rm eff}\equiv -(4\pi\hbar^2/m)(d^2 T(k)^{-1}/{dk^2})|_{k=0}$
which is determined explicitly from the scattering
model~Eq.~(\ref{eqTk})~\cite{Nota}.  Note this form in Eq.~(\ref{eqReff}) agrees
with the expression in~\cite{bruun}. The dimensionless width in units
of the Fermi energy is then the ratio $g\sqrt{n}/E_F$ as previously
discussed.

\begin{figure}[htbp]
\includegraphics[width=3.2in]{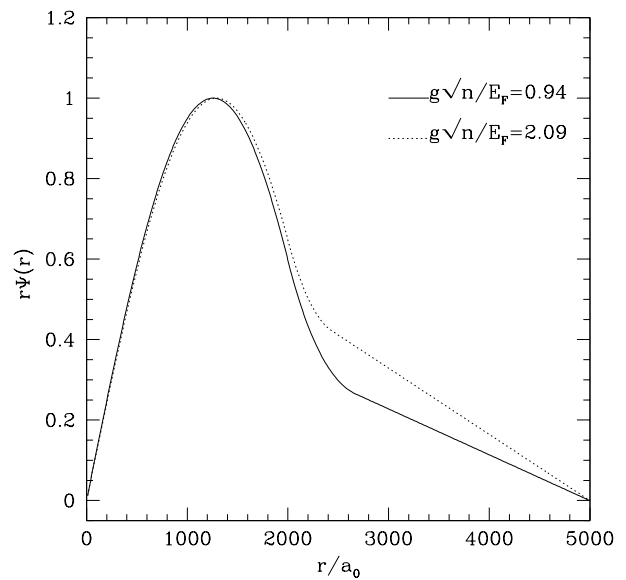}
\caption{The two-body wave function $r\Psi(r)$ corresponding to the 
case potentials $V(r)$ 1 and 8 in Fig.~\ref{fig:wb} and 
in Tab.~\ref{tab:cases}: $g\sqrt{n}/E_F=0.94$ (solid line), 
$2.09$ (dotted line).
The incoming kinetic energy is $E_k/k_B=1\; nK$. 
} \label{fig:psi2b}
\end{figure}

We list in the second column of Table~\ref{tab:cases} the set of
$g\sqrt{n}/E_F$ values that, together with the fixed condition
$a=+5000\; a_0$, define from here on our case studies. The
corresponding values of $r_1$ and $V_0$ are reported in the third and
fourth column.
\begin{table}[htbp]
\caption{Parameters for the model potential $V(r)$ in
Fig.~\ref{fig:wb}.  The values of $r_1$ and $V_0$ are reported in the
third and fourth column, that yield the different values of
$g\sqrt{n}/E_F$ in the second column with $a=5000a_0$ fixed. The
various cases are numbered as in the first column. 
The remaining model
parameters are $r_0=2000a_0$ and $V_1=100\mu K$.}
\bigskip
\begin{tabular}{l|l||l|l}
\colrule
Case& $g\sqrt{n}/E_F$ &$r_1/r_0$& $V_0 (\mu K)$\\ \colrule \colrule
1& 0.94& 1.3345 &45.0581\\ 2& 0.99& 1.3236 &44.9236\\ 3& 1.05& 1.312
&44.7673\\ 4& 1.12& 1.2994 &44.5806\\ 5& 1.32& 1.2712 &44.0898\\ 6&
1.48& 1.255 &43.7547\\ 7& 1.71& 1.237 &43.3292\\ 8& 2.09& 1.217
&42.7804\\ 9& 2.96& 1.1935 & 42.0159\\ 
\colrule \colrule
\end{tabular}
\label{tab:cases}
\end{table}
We illustrate in Fig.~\ref{fig:psi2b} the wavefunctions $r\Psi(r)$
that are found for the three $V(r)$ potentials reported in
Fig.~\ref{fig:wb} and labeled as cases $1$, $8$ and $10$ in
Table~\ref{tab:cases}, and for the incoming kinetic energy
$E_k/k_B=1\; nK$. While it is expected that the wavefunction should
vanish near $r=+5000\; a_0$, corresponding to the value of $a$, it is
clear that a knee develops in the probability amplitude for the
scattered wave in the spatial region within the barrier. This feature
becomes more pronounced as the resonance width decreases.

\section*{The nonlocal BCS equations}

We can now utilize our analysis of the two-body physics for this model
potential as an input to the many-body theory. In particular we wish
to find the ground-state of the Fermi gas interacting via $V(r)$. As
mentioned in the introduction, we aim in this paper to solve the
variational BCS scheme for the case of a nonlocal potential
interaction. Since our nonlocal potential allows us to explore the
consequences of scattering resonances, this is a suitable foundation
as eventual input to QMC schemes. Such schemes are necessary to
include the many-particle correlations which go beyond the pairing
fields studied here at the mean-field level in order to determine the
unitarity limit properties of the gas.

The ground state is determined within the variational BCS scheme by
means of the wave-function~\cite{bcs}:
\begin{equation}
|\Phi_0>= \Pi_{{\bf k}}(u_{{\bf k}}+v_{{\bf k}} a_{{\bf k},\uparrow}^
+ a_{-{\bf k},\downarrow}^{+})|0>\; ,
\label{wf}
\end{equation}
where $a_{{\bf k},\sigma}^+$ is the creation operator for electrons of
spin $\sigma$. The normalization of $\Phi_0$ leads to the condition
$|u_{{\bf k}}|^2+|v_{{\bf k}}|^2=1$.  The expected value of the
ground-state energy reads:
\begin{eqnarray}
E_0&=&\sum_\kvec 2\epsk|\vk|^2+\sum_{\kvec\qvec}\Vkq\uk\vk^*\uq\vq^*
\nonumber\\ &+& \sum_{\kvec\qvec}\Vkq\uk\uk^*\vq\vq^*
\label{eqEtot}
\end{eqnarray}
where $\epsk=\hbar^2{\bf k}^2/2m$. 

The summations in Eq.~(\ref{eqEtot}) can be converted into
one-dimensional integrals after performing the angular
integrations. For the generic function $F(\qvec)$, in our situation of
isotropic pairing this amounts to a substitution
\begin{equation}
\sum_\qvec \Vkq F(\qvec)\rightarrow \frac{1}{(2\pi)^3}\int dq q^2
V(k,q) F(q)\; ,
\label{equseful}
\end{equation}
with $V(k,q)$ determined from the three-dimensional Fourier transform 
of the spatial potential
\begin{eqnarray}
\label{eqVkq}
V(k,q)&=&\frac{2\pi}{kq} (V_0+V_1)\left[\frac{\sin
r_0|k+q|}{|k+q|}-\frac{\sin r_0|k-q|}{|k-q|}\right]\nonumber\\
&-&V_1\left[\frac{\sin r_1|k+q|}{|k+q|}-\frac{\sin
r_1|k-q|}{|k-q|}\right]\; .
\end{eqnarray}

The BCS solution is obtained by minimizing the free energy
$f=\bigl<\Phi_0|\hat{H}|\Phi_0\bigr>- \mu
\bigl<\Phi_0|\hat{N}|\Phi_0\bigr>$ with respect to the variational
parameters $u_k,v_k$. The chemical potential $\mu$ is determined by
the constraint to have the correct particle density.  The two
resulting equations to be self-consistently solved correspond
respectively to the isotropic superfluid gap and the particle density and are
given by
\begin{eqnarray}
\label{eqsDeltak}
\Delta(k)&=&
\frac{1}{(2\pi)^3}\int dq q^2 V(k,q) \frac{\Delta (q)}{2E(q)}\; \\
\label{eqsntot}
n&=&\frac{1}{(2\pi)^3} \int d\kvec\left(1-\frac{\xi (\kvec)}{E (\kvec)}\right)\;
\end{eqnarray}
In Eqs.~(\ref{eqsDeltak})-(\ref{eqsntot}), the excitation energy
$E(k)=\sqrt{\Delta (k)^2+\xi(k)^2}$ is expressed in terms of the gap
function and of the single-particle energy $\xi(\kvec)$
\begin{equation}
\label{eqsxik}
\xi(\kvec)=\epsk-\mu+\frac{1}{2}V_{k=q=0}n-\sum_\qvec
\Vkq\left(1-\frac{\xi(\qvec)}{E(\qvec)}\right)\; , 
\end{equation}
The second and third terms in Eq.~(\ref{eqsxik}) are the Hartree-Fock
corrections to the single-particle self-energy, that manifest as a
correction to the chemical potential.

We iteratively solve Eqs.~(\ref{eqsDeltak}) and (\ref{eqsntot}) until
self-consistency is obtained. Since two different length scales, $r_0$
and $k_F^{-1}$, are involved, integrations on Eqs.~(\ref{eqsDeltak})
and (\ref{eqsntot}) are executed by means of Gaussian integration in
three different grids of $k$-points that are in the ranges of $k_F$,
$r_0^{-1}$, and $+\infty$ (the latter is performed after changing to
the inverse variable $1/k$). We have typically used a total of
$N_k=1200$ grid points for this purpose.

The Bogoliubov $u_k$ and $v_k$ functions resulting from the energy
minimization are finally evaluated through the expressions
\begin{eqnarray}
\label{equkvk}
u_k&=&\mbox{sgn}(\Delta(k))
\sqrt{\frac{1}{2}\left(1+\frac{\xi(k)}{E(k)}\right)}\\ v_k&=&
\sqrt{\frac{1}{2}\left(1-\frac{\xi(k)}{E(k)}\right)}\; .
\end{eqnarray}

\section*{Self-consistent Results}

\begin{figure}[htb]
\includegraphics[width=3.2in]{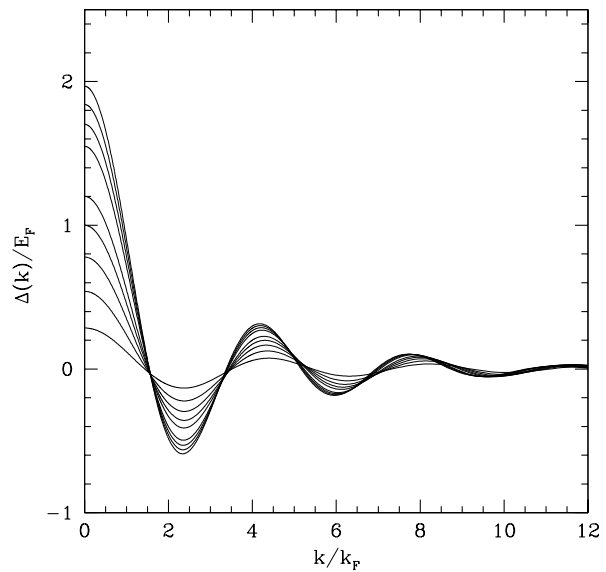}
\caption{$\Delta(k)$ in the unitarity limit with $a=5000\; a_0$, as a
function of the resonance width $g\sqrt{n}/E_F$. Curves with
increasing values of $\Delta(k=0)$ correspond to decreasing values of
$g\sqrt{n}/E_F$, from case $9$ to case $1$ in Table~\ref{tab:cases}.
}
\label{fig:Dk}
\end{figure}

The gap function $\Delta(k)$ is displayed in Fig.~\ref{fig:Dk} for
different values of $g\sqrt{n}/E_F$. The damped oscillatory behavior
on the scale of $1/r_0$ is a manifestation of the 
pairing potential, with a wavelength and damping coefficient
that are almost independent of the resonance width.

\begin{figure}[ht]
\includegraphics[width=3.2in]{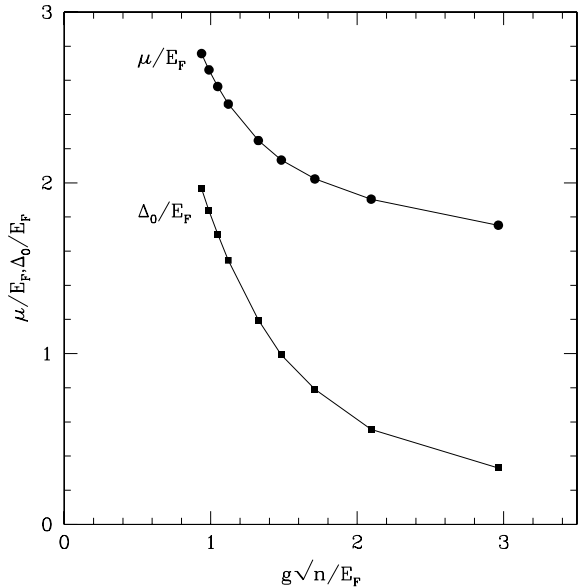}
\caption{$\Delta_0\equiv\Delta(k=0)$ (squares) and $\mu$ (circles) as
functions of $g\sqrt{n}/E_F$, with $a=+5000\; a_0$ fixed.  }
\label{fig:mu}
\end{figure}

The behavior of $\Delta_0\equiv\Delta(k=0)$ as a function of
$g\sqrt{n}/E_F$ is summarized by the squares 
in Fig.~\ref{fig:mu}, together with the behavior 
of the chemical potential (circles) following the gap.  
Fig.~\ref{fig:mu} demonstrates that a BCS-like solution emerges 
while the resonance shrinks on the scale of $E_F$, 
even on the BEC-side of the resonance, where the 
BCS-variational ansatz is
not expected to give a complete description. These results suggest 
that a high-quality resonance leads to a nonuniversal regime
in the unitarity limit. We have checked the consistency of this
picture by computing the self-consistent solutions
to Eqs.~(\ref{eqsDeltak})-(\ref{eqsntot}) in the case of narrower 
wells.
The narrowing of the resonance seems to increase the level of the interactions that are responsible for the superfluid pairing. This is
quantified in the ground-state energy (\ref{eqEtot}), whose values are
seen to become larger and negative with decreasing $g\sqrt{n}/E_F$
(see Tab.~\ref{tab:ene}).
\begin{table}[htbp]
\caption{Values of the total energy $E_0$(first column), chemical
  potential $\mu$ (second column) and of the gap
  $\Delta_0\equiv\Delta(k=0)$ at $k=0$ for the different values of
  $g\sqrt{n}/E_F$ in Tab.~\ref{tab:cases}.}
\bigskip
\begin{tabular}{l|l|l|l}
\colrule
 $g\sqrt{n}/E_F$ &$E_0/E_F$& $\mu/E_F$&$\Delta_0/k_F$\\
\colrule
\colrule
0.94&  -1.286760&   2.757061&   1.967685    \\
0.99&  -1.240605 &  2.661722 &  1.838010     \\
1.05&  -1.191324 &  2.563657 &  1.699060     \\
1.12&  -1.138285 &  2.461363 &  1.545724     \\
1.32&  -1.021072&   2.247637 &  1.195665    \\
1.48&  -0.953917 &  2.133496   &0.993021    \\
1.71&  -0.877986&   2.012829  & 0.771605   \\
2.09&  -0.816665  & 1.910900   &0.556854    \\
2.96&   -0.687847   &1.751602   &0.330776    \\
\colrule
\end{tabular}
\label{tab:ene}
\end{table}

\begin{figure}[ht]
\includegraphics[width=3.2in]{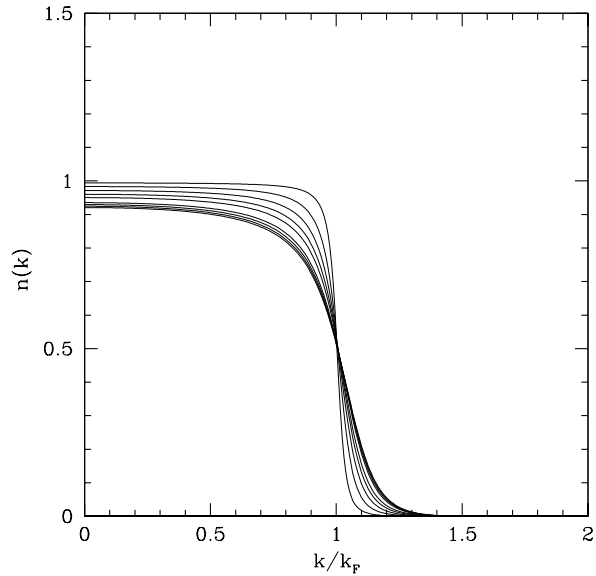}
\caption{Momentum distribution of the Fermi gas in the unitarity limit
with $a=5000\; a_0$, as a function of the resonance width
$g\sqrt{n}/E_F$. Curves with increasing values of $n(k=0)$ correspond
to increasing values of $g\sqrt{n}/E_F$, from case $1$ to case $9$ in
Tab.~\ref{tab:cases}.  } \label{fig:Nk}
\end{figure}

Further insight can be obtained from the analysis of the momentum
distribution $n(k)$
\begin{equation}
\label{eqmd}
n(k)=|v_k|^2\; , 
\end{equation}
that is reported in Fig.~\ref{fig:Nk}. In agreement with the
conclusions from Fig.~\ref{fig:mu}, larger values of $g\sqrt{n}/E_F$
correspond larger values of the jump at $k_F$ and thus to a normal
Fermi-gas character. The momentum distribution typical of a BCS
superfluid develops with decreasing values of $g\sqrt{n}/E_F$.

The superfluid state is also signaled by the emergence of a peak in
the pair distribution function, that is
\begin{equation}
\label{eqgr}
g(r)=g_{HF}(r)+g_p(r)\; ,
\end{equation}
where the Hartree-Fock contribution is
\begin{equation}
\label{eqgHF}
g_{HF}(r)=\frac{1}{4}-\left(\frac{1}{(2\pi)^3n}\right)^2 \int
d\kvec\, e^{-i\kvec\cdot\rvec}|\vk|^2
\end{equation}
and the pairing term 
\begin{equation}
\label{eqgp}
g_{p}(r)=+\left(\frac{1}{8\pi^3n}\right)^2 \int d\kvec\,
e^{-i\kvec\cdot\rvec}\uk\vk\; .
\end{equation}
The peak emerges on the scale of the interparticle distance
$r_sa_0$ with $r_s=(4\pi na_0^3/3)^{-1/3}$, 
and becomes better defined as long as the quality of the
resonant mode increases.
\begin{figure}[ht]
\includegraphics[width=3.2in]{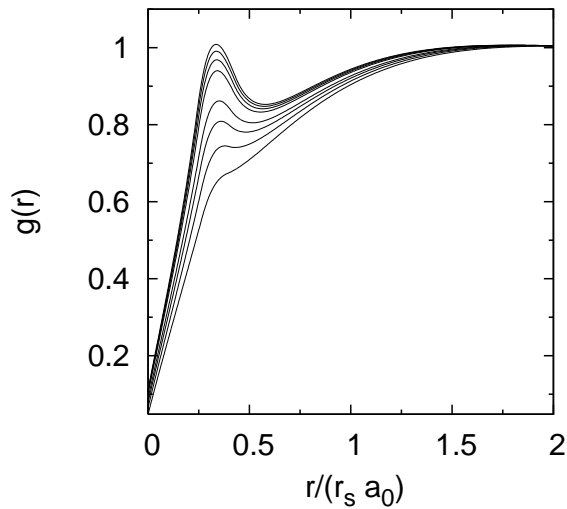}
\caption{Pair distribution function of the Fermi gas vs. the
  interparticle distance $r_sa_0$ in the unitarity limit with
  $a=5000\; a_0$.  Curves with increasing values of peak height
  correspond to decreasing values of $g\sqrt{n}/E_F$, from case $2$ to
  case $8$ in Tab.~\ref{tab:cases}.  } \label{fig:gr}
\end{figure}

\section*{Conclusions}

We have determined a solution to the many-body theory for the case of
a Feshbach resonance within the BCS-variational ansatz. The situation
we considered had a positive large scattering length corresponding to
a near unitarity limited system on the BEC side of the resonance. The
dependence of the scattering phase shift on energy was shown to be
important in the self-consistent solutions for the mean-fields.  The
solutions were found to depend sensitively on the resonance width. We
could vary all of these parameters within our model by modifying
the properties of an explicit and simple potential model consisting of
a potential well and potential barrier. Due to the fact that the
solution we obtained was for a nonlocal system, the resulting
many-body theory did not suffer from a formal ultraviolet divergence
and was therefore automatically renormalized.  The self-consistent
pairing field had to be determined at each wavevector value,
illuminating the nonlocal character of the solution.
The fact that we find a BCS solution for a positive value of
the scattering length can be understood from the total self-energy of
the system, which can be negative for narrow Feshbach resonances.

We emphasize that around the resonance, the BCS-variational ansatz is
not expected to give a complete description, although this is the
exclusively applied framework in which the BCS-BEC crossover theories
have been implemented to date. It is a necessary precursor to explore
and characterize the solutions we have found at the mean-field level
as inputs to a Quantum Monte Carlo numerical method,  
which is able to encapsulate
the many-particle correlations which we have dropped. This is an
important problem where it is anticipated that the unitarity limit of
the quantum system will emerge when the scattering resonance is broad,
but that a high-quality resonance will lead to a nonuniversal regime
which depends on microscopic resonance parameters.

M.~H. would like to acknowledge discussions 
with S. Giorgini and support from the National Science Foundation.
S.~D.~P. and M.~L.~C. would like to thank 
G. La Rocca for useful discussions. M.~L.~C. acknowledges support 
from the CRS-BEC of Trento, where has benefited from discussions with
S. Giorgini and S. Stringari, and from INFM through the project ATOMOPTICS.
S.~K. acknowledges support from the Netherlands Organisation for 
Scientific Research (NWO).


\begin{thebibliography}{xx}
\bibitem{Nota} Note that $R_{eff}$ always turns out to be negative
  close to resonance. In the present case with $a_{bg}=0$.
\end{thebibliography}
\end{document}